\documentclass[12pt]{article}
\usepackage{epsfig}
\usepackage{amsmath,amssymb}
\usepackage{graphicx}

\def\Reals{\mathop{\hbox{\mit I\kern-.2em R}}\nolimits}

\def\Complexes{{\hbox{\mit C\kern-.46em
            \vrule depth 0ex height 1.4ex width .05em\kern.41em}}}

\newtheorem{theorem}{Theorem}

\newtheorem{lemma}{Lemma}
\newtheorem{remark}{Remark}

\title{Leader-following Coordination of Multi-agent Systems with Coupling Time Delays}

\date{}

\author{Jiangping Hu, Yiguang Hong}

\begin{document}
\maketitle

\begin{center}
Key Lab of Systems and Control, Institute of Systems Science\\
Chinese Academy of Sciences, Beijing 100080, China
\end{center}

\begin{abstract}
In this paper, we consider a leader-following consensus problem of
a group of autonomous agents with time-varying coupling delays.
Two different cases of coupling topologies are investigated.  At
first, a necessary and sufficient condition is proved in the case
when the interconnection topology is fixed and directed. Then a
sufficient condition is proposed in the case when the coupling
topology is switched and balanced.  Numerical examples are also
given to illustrate our results.
\end{abstract}



\section{Introduction}
Recent years have witnessed steadily increasing recognition and
attention of coordinated motion of mobile agents across a broad
range of disciplines. Applications can be found in many areas such
as biology or ecology (e.g., aggregation behavior of animals in
\cite{amr, warb,bre}), physics (e.g., collective motion of
particles in \cite{vic95,vic00}), and engineering (e.g., formation
control of robots in \cite{lin05,hong, olf04,olf06}).  The studies
of multiple autonomous agents focus on understanding the general
mechanisms and interconnection rules of cooperative phenomena as
well as their potential applications in various engineering
problems.

In a multi-agent system, agents are usually coupled and
interconnected with some simple rules including nearest neighbor
rules \cite{vic95, olf04}. A computer graphics model to simulate
collective behavior of multiple agents was presented in
\cite{rey87}. With a proposed simple model and neighbor-based
rules, flocking and schooling were successfully simulated and
analyzed for self-propelled particles in \cite{vic95}. Also,
self-organized aggregation behavior of particle groups with
leaders becomes more and more interesting. The coordinated motion
of a group of motile particles with a leader has been analyzed in
\cite{mu}, while leader-follower networks have been also
considered in \cite{wang}.  Recently, to design distributed
flocking algorithms, Olfati-Saber has introduced a theoretical
framework including a virtual leader/follower architecture, which
is different from conventional leader/follower architecture
(\cite{olf06}).

Sometimes, the coupling delays between agents have to be taken
into consideration in practical problems (\cite{ear, koz, olf04}).
For example, \cite{ear} proposed a stability criterion for a
network of specific oscillators with time-delayed coupling. In
\cite{olf04}, the authors studied consensus problems of
continuous-time agents with interconnection communication delays.
The dynamics of each agent is first order and the graph to
describe the interconnection topology of these agents is
undirected.

In this paper, a leader-following consensus problem for multiple
agents with coupling time delays is discussed. Here the considered
dynamics of each agent is second order, coupling time delay is
time-varying, and the interconnection graph of the agents is
directed. The convergence analysis of the consensus problem with
directed graphs (or digraph for short) is more challenging than
that of undirected graphs due to the complexity of directed
graphs. The analysis becomes harder if time delay is involved. For
time-delay systems, modeled by delayed differential equations, an
effective way to deal with convergence and stability problems is
Lyapunov-based; Lyapunov-Krasovskii functionals or
Lyapunov-Razumikhin functions are often used in the analysis
\cite{hale}.

The paper is organized as follows.  Section 2 presents the
multi-agent model and some preliminaries. Then, two cases, fixed
coupling topology and switched coupling topology, are considered.
The leader-following convergence of two models in the two cases
are analyzed in Section 3 and Section 4, respectively. Here,
Lyapunov-Razumikhin functions are employed, along with the
analysis of linear matrix inequalities. Finally, some concluding
remarks are given in Section 5.

By convention, $R$ and $Z^{+}$ represent the real number set and
the positive integer set, respectively; $I_n$ is an $n\times n $
identity matrix; for any vector $x$, $x^{T}$ denotes is its
transpose; $||\cdot||$ denotes Euclidean norm.

\section{Model Description}

We consider a group of $n+1$ identical agents, in which an agent
indexed by 0 is assigned as the ``leader" and the other agents
indexed by $1,...,n$ are referred to as ``follower-agents" (or
``agents" when no confusion arises). The motion of the leader is
independent and the motion of each follower is influenced by the
leader and the other followers. A continuous-time model of the $n$
agents is described as follows:
\begin{equation}
\label{move1} \ddot {x}_i = u_i,\quad i=1,...,n,
\end{equation}
or equivalently,
\begin{equation}
\label{move2} \left\{ {\begin{array}{l}
 \dot {x}_i = v_i, \\
 \dot {v}_i = u_i, \\
 \end{array}} \right.
\end{equation}
where the state $x_i\in R^{m}$ can be the position vector of agent
$i$, $v_i\in R^{m}$ its velocity vector and $u_i\in R^{m}$ its
coupling inputs for $i=1,...,n$. Denote
\begin{equation*}
x=\begin{pmatrix} x_1 \\ x_2 \\ \vdots \\x_n \end{pmatrix},\quad
v=\begin{pmatrix} v_1 \\ v_2 \\ \vdots \\v_n \end{pmatrix}, \quad
u=\begin{pmatrix} u_1 \\ u_2 \\ \vdots \\u_n \end{pmatrix} \in
R^{mn}. \end{equation*}

Without loss of generality, in the study of leader-following
stability, we take $m=1$ for simplicity in the sequel. Then
(\ref{move2}) can be rewritten as
\begin{equation}
\label{move3} \left\{ {\begin{array}{l}
 \dot {x} = v,\\
 \dot {v} = u \in R^n.\\
 \end{array}} \right.
\end{equation}
The dynamics of the leader is described as follows:
\begin{equation}
\label{moveleader} \dot x_0=v_0\in R,
\end{equation}
where $v_0$ is the desired constant velocity.

If each agent is regarded as a node, then their coupling topology
is conveniently described by a simple graph (basic concepts and
notations of graph theory can be found in \cite{bang, god,
olf04}).  Let $\mathcal{G}=(\mathcal{V},\mathcal{E},A)$ be a
weighted digraph of order $n$ with the set of nodes
$\mathcal{V}=\{1,2,...,n\}$, set of arcs $\mathcal{E}\subseteq
\mathcal{V} \times \mathcal{V}$, and a weighted adjacency matrix
$A=[a_{ij}]\in R^{n \times n}$ with nonnegative elements. The node
indexes belong to a finite index set $\mathcal{I}=\{1,2,...,n\}$.
An \emph{arc} of $\mathcal{G}$ is denoted by $(i,j)$, which starts
from $i$ and ends on $j$. The element $a_{ij}$ associated with the
arc of the digraph is positive, i.e. $a_{ij}>0 \Leftrightarrow
(i,j) \in \mathcal{E}$. Moreover, we assume $a_{ii}=0$ for all $i
\in \mathcal{I}$. The set of neighbors of node $i$ is denoted by
$\mathcal{N}_{i}=\{j\in \mathcal{V}:(i,j)\in \mathcal{E}\}$. A
\emph{cluster} is any subset $\mathcal{J}\subset \mathcal{V}$ of
the nodes of the digraph. The set of neighbors of a cluster
$\mathcal{J}$ is defined by
$\mathcal{N}_{\mathcal{J}}=\bigcup_{i\in
\mathcal{J}}\mathcal{N}_{i}=\{j\in \mathcal{V}:i\in
\mathcal{J},(i,j)\in \mathcal{E}\}$. A \emph{path} in a digraph is
a sequence $i_0,i_1,\cdots,i_f$ of distinct nodes such that
$(i_{j-1},i_j)$ is an arc for $j=1,2,\cdots,f,f\in Z^+$. If there
exists a path from node $i$ to node $j$, we say that $j$ is
reachable from $i$. A digraph $\mathcal{G}$ is strongly connected
if there exists a path between any two distinct nodes. A
\emph{strong component} of a digraph is an induced subgraph that
is maximal, subject to being strongly connected. Moreover, if
$\sum_{j \in \mathcal{N}_{i}}a_{ij}=\sum_{j \in
\mathcal{N}_{i}}a_{ji}$ for all $i=1,...,n$, then the digraph
$\mathcal{G}$ is called \emph{balanced}, which was first
introduced in \cite{olf04}.

A diagonal matrix $D=diag\{ d_1,...,d_n\}\in R^{n\times n}$ is a
degree matrix of $\mathcal{G}$, whose diagonal elements
$d_i=\sum_{j \in \mathcal{N}_{i}}a_{ij}$ for $i=1,...,n$. Then the
Laplacian of the weighted digraph $\mathcal{G}$ is defined as
\begin{equation}
 L=D-A\in R^{n\times n}.
\end{equation}

To study a leader-following problem, we also concern another graph
$\bar{\mathcal{G}}$ associated with the system consisting of $n$
agents and one leader (labelled $0$).  Similarly, we define a
diagonal matrix $B\in R^{n\times n}$ to be a leader adjacency
matrix associated with $\bar{\mathcal{G}}$ with diagonal elements
$b_i\;(i\in \mathcal{I})$, where $b_i=a_{i0}$ for some constant
$a_{i0}>0$ if node $0$ (i.e., the leader) is a neighbor of node
$i$ and $b_i=0$ otherwise. For $\bar{ \mathcal{G}}$, if there is a
path in $\bar {\mathcal{G}}$ from every node $i$ in $\mathcal{G}$
to node $0$, we say that node $0$ is globally reachable in $\bar{
\mathcal{G}}$, which is much weaker than strong connectedness.

{\bf Example 1}.  As shown in Figs. 1 and 2, both
$\bar{\mathcal{G}}_1$ and $\bar{\mathcal{G}}_2$ are not strongly
connected, but they have a globally reachable node $0$. Suppose
that the weight of each arc is $1$ in both cases. Obviously,
$\mathcal{G}_2$ with $\mathcal{V}=\{1,2,3,4\}$ is balanced.

Laplacians of $\mathcal{G}_1$ and $\mathcal{G}_2$ as well as the
leader adjacency matrices $B_1,\;B_2$ are easily obtained as
follows:
\begin{align*}
L_1=\left(\begin{array}{cccc}
1&-1&0&0\\-1&1&0&0\\0&0&0&0\\0&-1&-1&2
\end{array}\right), L_2=\left(\begin{array}{cccc}
1&-1&0&0\\-1&1&0&0\\0&0&1&-1\\0&0&-1&1
\end{array}\right), B_1=B_2=\left(\begin{array}{cccc}
1&0&0&0\\0&0&0&0\\0&0&1&0\\0&0&0&0
\end{array}\right).
\end{align*}

\begin{center}{{\includegraphics[scale=0.2]{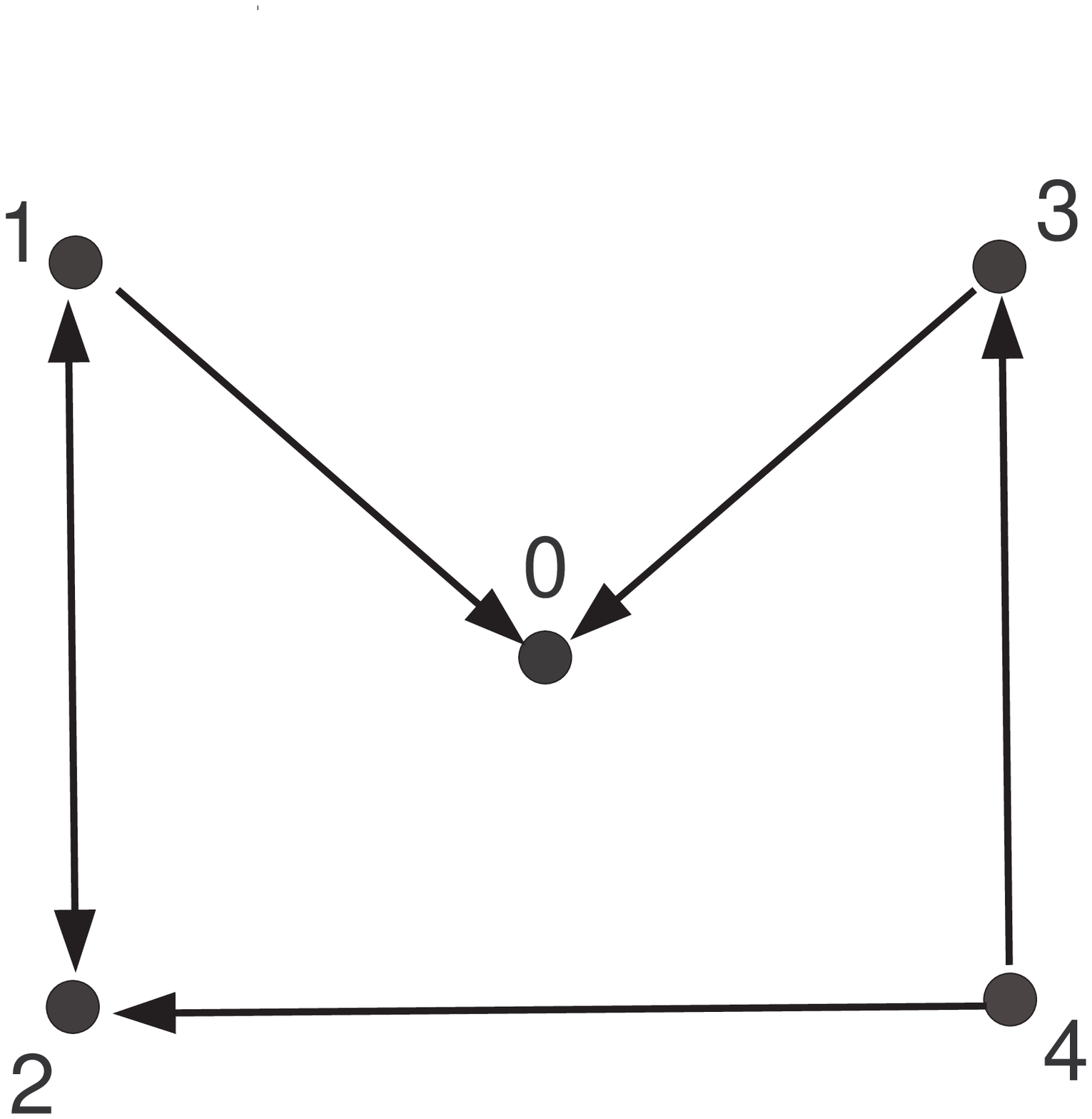}
}\hskip 1.5cm {\includegraphics[scale=0.2]{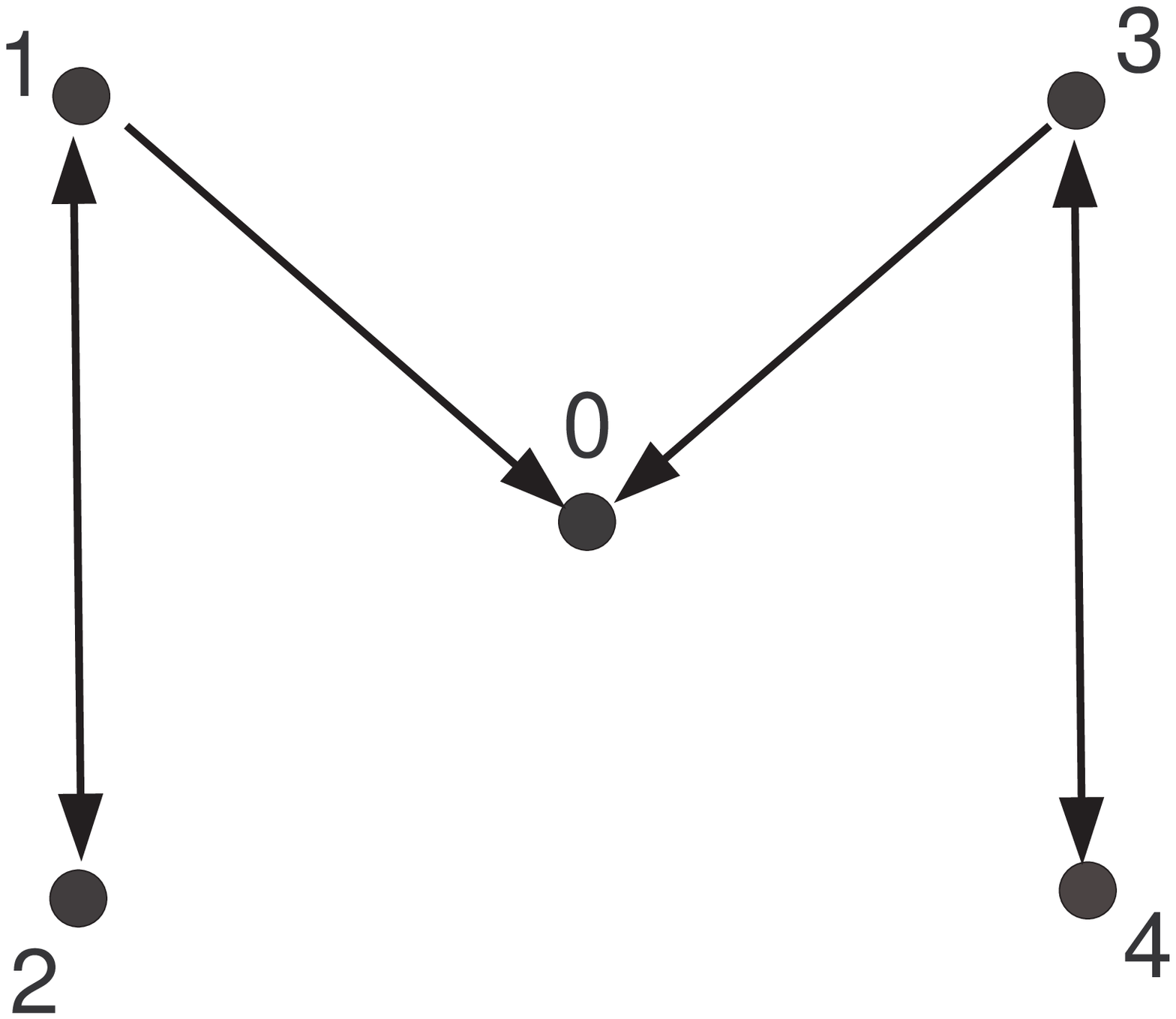} }

{Fig.1 $\bar{\mathcal{G}}_1$ and $\mathcal{G}_1$\hskip 2.6cm Fig.2
$\bar{\mathcal{G}}_2$ and $\mathcal{G}_2$}}
\end{center}

The following lemma was obtained in (\cite{lin05, more}).

\begin{lemma}
\label{lemgr1} A digraph $\mathcal{G}=(\mathcal{V},\mathcal{E},A)$
has a globally reachable node if and only if for every pair of
nonempty, disjoint subsets $\mathcal{V}_1,\mathcal{V}_2\subset
\mathcal{V}$ satisfies
$\mathcal{N}_{S_i}\bigcup\mathcal{N}_{S_j}\neq \emptyset$.
\end{lemma}

\begin{remark}
Let $S_1,S_2,...,S_p$ be the strong components of $\mathcal{G}=
(\mathcal{V}, \mathcal{E})$ and $\mathcal{N}_{S_i}$ be the
neighbor sets for $S_i,i=1,...,p, p> 1$. From Lemma \ref{lemgr1},
a digraph $\mathcal{G} $  has a globally reachable node if and
only if every pair of $S_i,S_j$ satisfies
$\mathcal{N}_{S_i}\bigcup\mathcal{N}_{S_j}\neq \emptyset$. If the
graph is strongly connected, then each node is globally reachable
from every other node.
\end{remark}

The next lemma shows an important property of Laplacian $L$
(\cite{lin05}).

\begin{lemma}
\label{lemgr2} The digraph $\mathcal{G}$ has a globally reachable
node if and only if Laplacian $L$ of $\mathcal{G}$ has a simple
zero eigenvalue (with eigenvector $\textbf{1}=(1,...,1)^{T}\in
R^{n}$).
\end{lemma}

Due to the coupling delays, each agent cannot instantly get the
information from others or the leader. Thus, for agent
$i\;(i=1,...,n)$, a neighbor-based coupling rule can be expressed
as follows:
\begin{equation}
\label{prot} u_i(t)=\sum_{j \in
\mathcal{N}_{i}(\sigma)}a_{ij}(x_j(t-r)-x_i(t-r))
+b_i(\sigma)(x_0(t-r)-x_i(t-r))+k(v_0-v_i(t)),\; k>0,
\end{equation}
where the time-varying delay $r(t)>0$ is a continuously
differentiable function with
\begin{equation}
\label{const1} 0<r<\tau,
\end{equation}
$\sigma: [0,\infty) \to \mathcal{I}_{\Gamma}=\{1,...,N\}$ ($N$
denotes the total number of all possible digraphs) is a switching
signal that determines the coupling topology. The set
$\Gamma=\{\mathcal{G}_1,...,\mathcal{G}_N\}$ is a finite
collection of graphs with a common node set $\mathcal{V}$. If
$\sigma$ is a constant function, then the corresponding
interconnection topology is fixed. In addition,
$\mathcal{N}_{i}(\sigma)$ is the index set of neighbors of agent
$i$ in the digraph $\mathcal{G}_{\sigma}$ while
$a_{ij}\;(i,j=1,...,n)$ are elements of the adjacency matrix of
$\mathcal{G}_{\sigma}$ and $b_i(\sigma)\;(i=1,...,n)$ are the
diagonal elements of the leader adjacency matrix  associated with
 $\bar{\mathcal{G}}_\sigma$.

With (\ref{prot}), (\ref{move2}) can be written in a matrix form:
\begin{equation}
\label{model1}
\begin{cases}
 \dot{x}=v,\\
\dot{v}=-(L_\sigma+B_\sigma)x(t-r)-k(v-v_0\mathbf{1})+B_\sigma\mathbf{1}x_0(t-r),
\end{cases}
\end{equation}
where $L_\sigma$ is Laplacian of $\mathcal{G}_\sigma$ and
$B_\sigma$ is the leader adjacency matrix associated with
 $\bar{\mathcal{G}}_\sigma$.

In the sequel, we will demonstrate the convergence of the dynamics
system (\ref{model1}); that is, $x_i \to x_0,v_i \to v_0$ as $t
\to \infty$.

\section{Fixed Coupling Topology}

In this section, we will focus on the convergence analysis of a
group of dynamic agents with fixed interconnection topology.  In
this case, the subscript $\sigma$ can be dropped.

Let $\bar{x}=x-x_0\mathbf{1},\bar{v}=v-v_0\mathbf{1}$. Because
$-(L+B)x(t-r)+B\textbf{1}x_{0}(t-r) =-(L+B)\bar x(t-r)$ (invoking
Lemma \ref{lemgr2}), we can rewrite system (\ref{model1}) as
\begin{equation}
\label{model2} \dot{\epsilon}=C\epsilon(t)+E\epsilon(t-r),
\end{equation}
where
$$
\epsilon=\left(\begin{array}{c} \bar x\\
\bar v\end{array}\right),\; C=\begin{pmatrix} 0&I_n\\
0&-kI_n
\end{pmatrix},\; E=\begin{pmatrix} 0&0\\
-H&0
\end{pmatrix},\; H=L+B.
$$

Before the discussion, we introduce some basic concepts or results
for time-delay systems (\cite{hale}). Consider the following
system:
\begin{equation}
\begin{cases}
\label{delay}\dot x=f(x_t),\quad t > 0 ,\\
x(\theta)=\varphi(\theta),\;\theta \in [-\tau,0],
\end{cases}
\end{equation}
where $x_t(\theta)=x(t+\theta),\forall \theta\in [-\tau,0]$ and
$f(0)=0$. Let $C([-\tau,0],R^{n})$ be a Banach space of continuous
functions defined on an interval $[-\tau, 0]$, taking values in
$R^{n}$ with the topology of uniform convergence, and with a norm
$||\varphi||_c = \max\limits_{\theta \in [-\tau,
0]}||\varphi(\theta)||$. The following result is for the stability
of system (\ref{delay}) (the details can be found in \cite{hale}).

\begin{lemma}
\label{lem1}(Lyapunov-Razumikhin Theorem) Let $\phi_1,\phi_2$, and
$\phi_3$ be continuous, nonnegative, nondecreasing functions with
$\phi_1(s)>0,\phi_2(s)>0,\phi_3(s)>0$ for $s >0$ and
$\phi_1(0)=\phi_2(0)=0$. For system (\ref{delay}), suppose that
the function $f: C([-\tau,0],R^{n}) \to R$ takes bounded sets of
$C([-\tau,0],R^{n})$ in bounded sets of $R^n$. If there is a
continuous function $V(t,x)$ such that
\begin{equation}
 \label{cond1}\phi_1(||x||) \leq  V(t,x) \leq
 \phi_2(||x||),\;t\in R, \;x\in R^{n}.
 \end{equation}
In addition, there exists a continuous nondecreasing function
$\phi(s)$ with $\phi(s)>s,\; s>0$ such that
\begin{equation}
\label{cond2} \dot{V}(t,x)|_{(\ref{delay})} \leq
-\phi_3(||x||),\quad \mbox{if}\;\;
V(t+\theta,x(t+\theta))<\phi(V(t,x(t))),\;\theta \in [-\tau,0],
 \end{equation}
 then the solution $x=0$ is uniformly asymptotically stable.
\end{lemma}

Usually, $V(t,x)$ is called Lyapunov-Razumikhin function if it
satisfies (\ref{cond1}) and (\ref{cond2}) in Lemma \ref{lem1}.

\begin{remark}
\label{rem2} Lyapunov-Razumikhin theorem indicates that it is
unnecessary to require that $\dot V(t,x)$ be non-positive for all
initial data in order to have stability of system (\ref{delay}).
In fact, one only needs to consider the initial data if a
trajectory of equation (\ref{delay}) starting from these initial
data is ``diverging" (that is, $V(t+\theta, x(t+\theta))<
\phi(V(t,x(t)))$ for all $\theta \in [-\tau,0]$ in (\ref{cond2})).
\end{remark}

A matrix $A$ is said to have \emph{property SC} (\cite{horn}) if,
for every pair of distinct integers $\hbar,\ell$ with $1 \leq
\hbar,\ell \leq n$, there is a sequence of distinct integers
$\hbar=i_1,i_2,...,i_{j-1}, i_j=\ell,1 \leq j \leq n$ such that
all of the matrix entries
$a_{i_1i_2},a_{i_2i_3},...,a_{i_{j-1}i_j}$ are nonzero. In fact,
it is obvious that, if $\mathcal{G}$ is strongly connected, then
its adjacency matrix $A$ has \emph{property SC}.  Moreover, a
matrix is called a positive stable matrix if its eigenvalues have
positive real-parts. Note that $H=L+B$ plays a key role in the
convergence analysis of system (\ref{model2}). The following lemma
shows a relationship between $H$ and the connectedness of graph
$\bar {\mathcal{G}}$ (as defined in Section 2).

\begin{lemma}
\label{lemgr3} The matrix $H=L+B$ is positive stable if and only
if node $0$ is globally reachable in $\bar {\mathcal{G}}$.
\end{lemma}

Proof: (Sufficiency) Based on $Ger\check{s}gorin$ disk theorem
(\cite{horn}), all the eigenvalues of $H$ are located in the union
of $n$ discs:
\begin{equation*}
Ger(H)=\bigcup_{i=1}^{n}\{z\in R^{2}:|z-d_{i}-b_i| \leq \sum_{j
\neq i}a_{ij}\}.
\end{equation*}
However, for the graph $\mathcal{G}$, $d_{i}=\sum_{j\neq
i}a_{ij}$. Thus, every disc with radius $d_i$ will be located in
the right half of the complex plane, and then $H$ has either zero
eigenvalue or eigenvalue with positive real-part. Since node $0$
is globally reachable, there exists at least one $b_i>0$.
Therefore, at least one $Ger\check{s}gorin$ circle does not pass
through the origin.

The following two cases are considered to prove the sufficient
condition:
\begin{description}
\item Case (i) {\it $\mathcal{G}$ has a globally reachable node}:
Let $S_1,...,S_p$ ($p\in Z^+$) be the strong components of
$\mathcal{G}$. If $p=1$, $\mathcal{G}$ is strongly connected. Then
its adjacency matrix $A$ has \emph{property SC}. Since $D+B$ is a
diagonal matrix with nonnegative diagonal entries, $H$ still has
\emph{property SC}. By Better theorem (\cite{horn}), if zero is an
eigenvalue of $H$, it is just a boundary point of $Ger(H)$.
Therefore, every $Ger\check{s}gorin$ circle passes through zero,
which leads to a contradiction. Hence, zero is not an eigenvalue
of $H$.

If $p>1$, then there is one strong component, say $S_1$, having no
neighbor set by Lemma \ref{lemgr1}.  We rearrange the indices of
$n$ agents such that the Laplacian of $\mathcal{G}$ is taken in
the form
\begin{equation}
\label{lap00} L=\begin{pmatrix}L_{11} & 0\\L_{21}&
L_{22}\end{pmatrix},
\end{equation}
where $L_{11}\in R^{\kappa\times \kappa}\;(\kappa <n)$ is
Laplacian of the component $S_1$. From Lemma \ref{lemgr2}, zero is
a simple eigenvalue of $L_{11}$ and $L$, while $L_{22}$ is
nonsingular. Since node $0$ is globally reachable, then the block
matrix $B_1\neq 0$ with $B=diag\{B_1,B_2\}$. Similar to the case
when $p=1$, we conclude that zero is not an eigenvalue of
$L_{11}+B_1$, and is also not an eigenvalue of $H$.

\item Case (ii) {\it $\mathcal{G}$ has no globally reachable
node}: Let $S_1,...,S_p$ be the strong components with
$\mathcal{N}_{S_i}=\emptyset,i=1,...,p, p>1$ by Lemma
\ref{lemgr1}. Since $\bigcup_{i=1}^{p}\mathcal{V}(S_i) \subset
\mathcal{V}(\mathcal{G})$, Laplacian associated with $\mathcal{G}$
can be transformed to the following form:
\begin{equation}
\label{lap02}
L=\left(\begin{array}{cccc} L_{11}&&&\\
&\ddots&&\\
&&L_{pp}&\\
L_{p+1,1}&\cdots&L_{p+1,p}&L_{p+1,p+1}
\end{array}\right),
\end{equation}
where $L_{ii}$ is the Laplacian associated with $S_i$ for
$i=1,...,p$. One can easily verify that $L_{p+1,p+1}$ is
nonsingular. Since node $0$ is globally reachable, then $B_i\neq
0$ for $i=1,...,p$ where $B_i$, corresponding to $L_{ii}$, are
diagonal blocks of $B$. Similar to the proof in Case (i), we can
obtain that zero is not eigenvalue of $H_i$ or $H$.
\end{description}

(Necessity) If node $0$ is not globally reachable in
$\bar{\mathcal{G}}$, then we also have:
\begin{description}
\item Case (i) {\it $\mathcal{G}$ has a globally reachable node}:
As discussed before, assume $S_1$ has no neighbor set, and then we
have (\ref{lap00}), where $L_{11}\in R^{\kappa\times \kappa}\;
(\kappa\in Z^+)$ is the Laplacian of $S_1$. Invoking Lemma
\ref{lemgr2}, zero is a simple eigenvalue of $L_{11}$ and $L$,
while $L_{22}$ is nonsingular. By the assumption that node $0$ is
not globally reachable in $\bar{\mathcal{G}}$, then the block
matrix $B_1=0$ with $B=diag\{B_1,B_2\}$. Therefore, zero is a
simple eigenvalue of $L_{11}+B_1$, and is also a simple eigenvalue
of $H$.  This leads to a contradiction.

\item Case (ii) {\it $\mathcal{G}$ has no globally reachable
node}: As discussed before, we have (\ref{lap02}). By the
assumption that node $0$ is not globally reachable in
$\bar{\mathcal{G}}$, then there exists at least one $B_i=0$ for
$i=1,...,p$ where $B_i$, corresponding to $L_{ii}$, are diagonal
blocks of $B$. Thus, $H_i$ and $H$ have more than one zero
eigenvalues.  This implies a contradiction.
\end{description}\hfill\rule{4pt}{8pt}

Therefore, if node $0$ is globally reachable in
$\bar{\mathcal{G}}$, $H$ is positive stable, and from Lyapunov
theorem, there exists a positive definite matrix $\bar P\in
R^{n\times n}$ such that
\begin{equation}
\label{lyaequ}\bar P H+H^{T}\bar P=I_n.
\end{equation}

Let $\bar\mu=max\{\mbox{eigenvalues of} \;\bar P HH^T\bar P\}$ and
let $\bar \lambda$ be the smallest eigenvalue of $\bar P$. Now we
give the main result as follows.

\begin{theorem}
\label{thm1} For system (\ref{model2}), take
\begin{equation}
\label{kstar} k>k^*=\frac{\bar{\mu}}{2\bar\lambda}+1.
\end{equation}
Then, when $\tau$ is sufficiently small,
\begin{equation}
\label{conclu}\lim_{t \to \infty}\epsilon(t)=0,
\end{equation}
if and only if node $0$ is globally reachable in
$\bar{\mathcal{G}}$.
\end{theorem}

Proof: (Sufficiency) Since node $0$ is globally reachable in
$\bar{\mathcal{G}}$, $H$ is positive stable and $\bar P$ is a
positive definite matrix satisfying (\ref{lyaequ}). Take a
Lyapunov-Razumikhin function $V(\epsilon)=\epsilon^{T}P\epsilon$,
where
$$
P=\left(\begin{array}{cc}k\bar P&\bar P\\\bar P&\bar P
\end{array}\right)\qquad(k>1)
$$
is positive definite.

Then we consider $\dot V(\epsilon)|_{(\ref{model2})}$.

By \emph{Leibniz-Newton} formula,
\begin{align*}
\epsilon(t-r)&=\epsilon(t)-\int_{-r}^{0}\dot \epsilon(t+s)ds\\
&=\epsilon(t)-C\int_{-r}^{0} \epsilon(t+s)ds-E\int_{-2r}^{-r}
\epsilon(t+s)ds.
\end{align*}
Thus, from $E^2=0$, the delayed differential equation
(\ref{model2}) can be rewritten as
\begin{equation*}
\dot \epsilon=F\epsilon-EC\int_{-r}^{0}\epsilon(t+s)ds,
\end{equation*}
where $F=C+E$.

Note that $2a^{T}b \leq a^{T}\Psi a+b^{T}\Psi^{-1}b$ holds for any
appropriate positive definite matrix $\Psi$.  Then, with
$a=-C^TE^TP\epsilon,b= \epsilon(t+s)$ and $\Psi=P^{-1}$, we have
$$
\dot{V}|_{(\ref{model2})}=\epsilon^{T}(F^{T}P+PF)
\epsilon-2\epsilon^TPEC\int_{-r}^{0}
 \epsilon(t+s)ds
$$
$$
 \leq \epsilon^{T}(F^{T}P+PF)
\epsilon+r\epsilon^TPECP^{-1}C^TE^TP\epsilon+\int_{-r}^{0}
 \epsilon^T(t+s)P\epsilon(t+s)ds.
$$

Take $\phi(s)=qs$ for some constant $q>1$.  In the case of
\begin{equation}
\label{cond3}V(\epsilon(t+\theta))<qV(\epsilon(t)),\;-\tau\leq
\theta \leq 0,
\end{equation}
we have
$$
\dot{V}
\leq-\epsilon^{T}Q\epsilon+r\epsilon^T(PECP^{-1}C^TE^TP+qP)\epsilon,
$$
where
$$
Q=-(F^TP+PF)=\begin{pmatrix}I_n&H^T\bar P\\\bar PH&2(k-1)\bar
P\end{pmatrix}.
$$
$Q$ is positive definite if $k$ satisfies (\ref{kstar}), according
to Lemma \ref{lemgr3} and Schur complements theorem (\cite{horn}).
Let $\lambda_{min}$ denote the minimum eigenvalues of $Q$. If we
take
\begin{equation}
\label{bound1}
r<\tau=\frac{\lambda_{min}}{||PECP^{-1}C^TE^TP||+q||P||},
\end{equation}
then $\dot V(\epsilon)\leq
-\eta \epsilon^T\epsilon$ for some $\eta>0$.  Therefore, the
conclusion follows by Lemma \ref{lem1}.

(Necessity) Since system (\ref{model2}) is asymptotically stable,
the eigenvalues of $F$ have negative real-parts, which implies
that $H$ is positive stable. By Lemma \ref{lemgr3}, node $0$ is
globally reachable in $\bar{\mathcal{G}}$. \hfill\rule{4pt}{8pt}

\begin{remark}
\label{rem3}  In the proof of Theorem \ref{thm1}, we have obtained
a finite bound of the considered time-varying delay, that is,
$\tau$ in (\ref{bound1}), though ``$\tau$ is sufficiently small"
is mentioned in Theorem \ref{thm1}.
\end{remark}

\begin{remark}
Obviously, (\ref{conclu}) still holds if the time delay is
constant. Moreover, if the system (\ref{move2}) is free of
time-delay (that is, $r\equiv 0$), then the coupling rule
(\ref{prot}) becomes
$$
u_i(t)=\sum_{j \in
\mathcal{N}_{i}(\sigma)}a_{ij}(x_j(t)-x_i(t))+b_i(\sigma)(x_0(t)-x_i(t))+k(v_0-v_i(t)),
$$
which is consistent with the nearest neighbor rules in
\cite{olf04}.
\end{remark}

For illustration, we give an numerical example with the
interconnection graph given in Fig. 1.  It is not hard to obtain
$$
\bar \mu =0.3139,\;\bar \lambda =0.1835,\;k^*=2.7106,
$$
$$
\lambda_{min}=0.3325,\;q=1.0500,\;\tau=0.0334,
$$
$$
\bar P=\left(\begin{array}{cccc}
 0.5379 &   0.5758 &   0.0439  &  0.0227\\
 0.5758 &   1.1667 &   0.1091 &   0.0909\\
 0.0439 &   0.1091 &   0.5833 &   0.0833\\
 0.0227 &   0.0909 &   0.0833 &   0.2500
\end{array}\right).
$$
Take $k=3$ and the time-varying delay $r(t)=0.0300|\cos(t)|$ in
the simulation.

Fig. 3 shows the simulation results for both position errors and
velocity errors, while Fig. 4 demonstrates that the trajectories
of the four agents and the one of the leader.

\begin{center}{\includegraphics[scale=0.4]{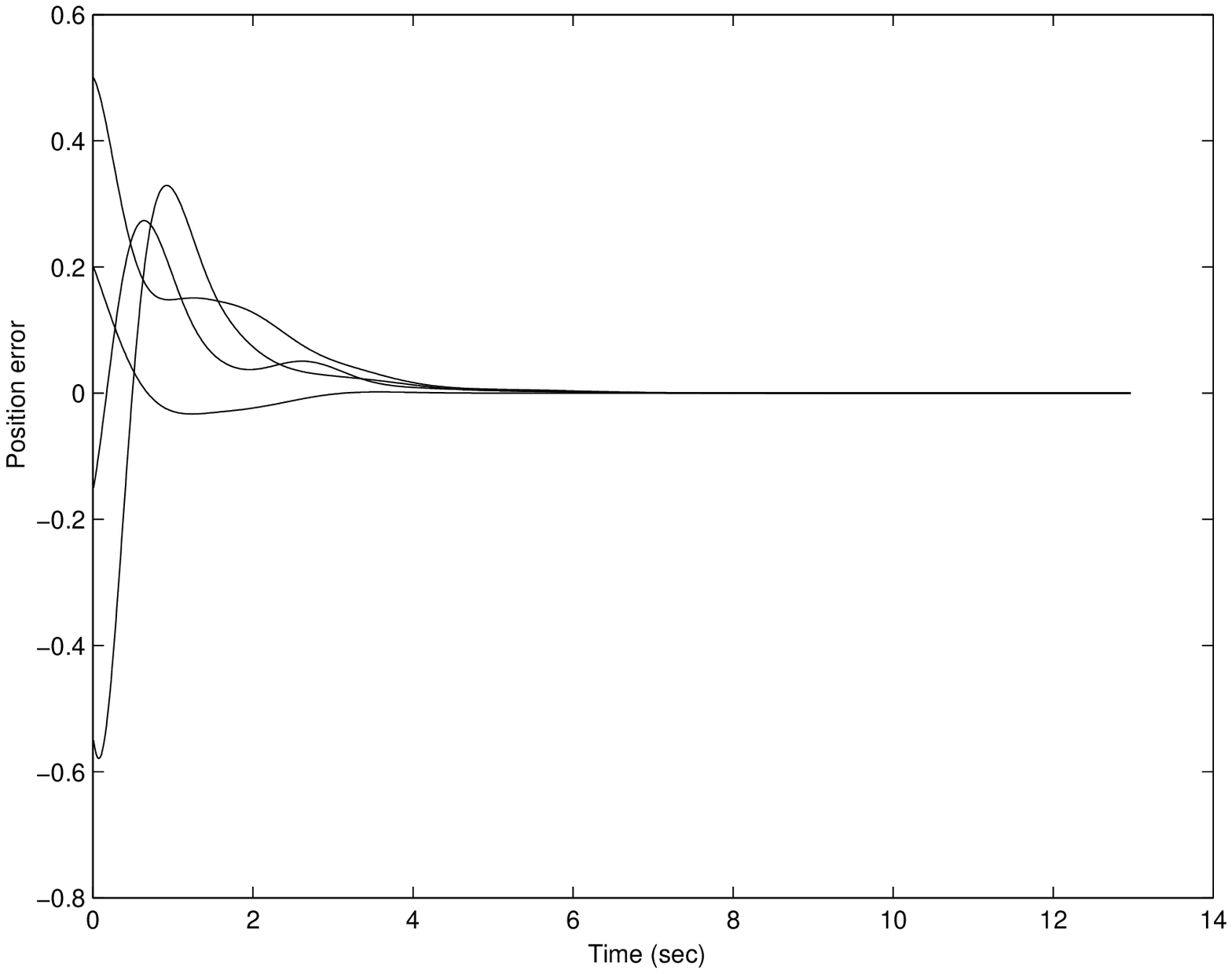}\includegraphics[scale=0.4]{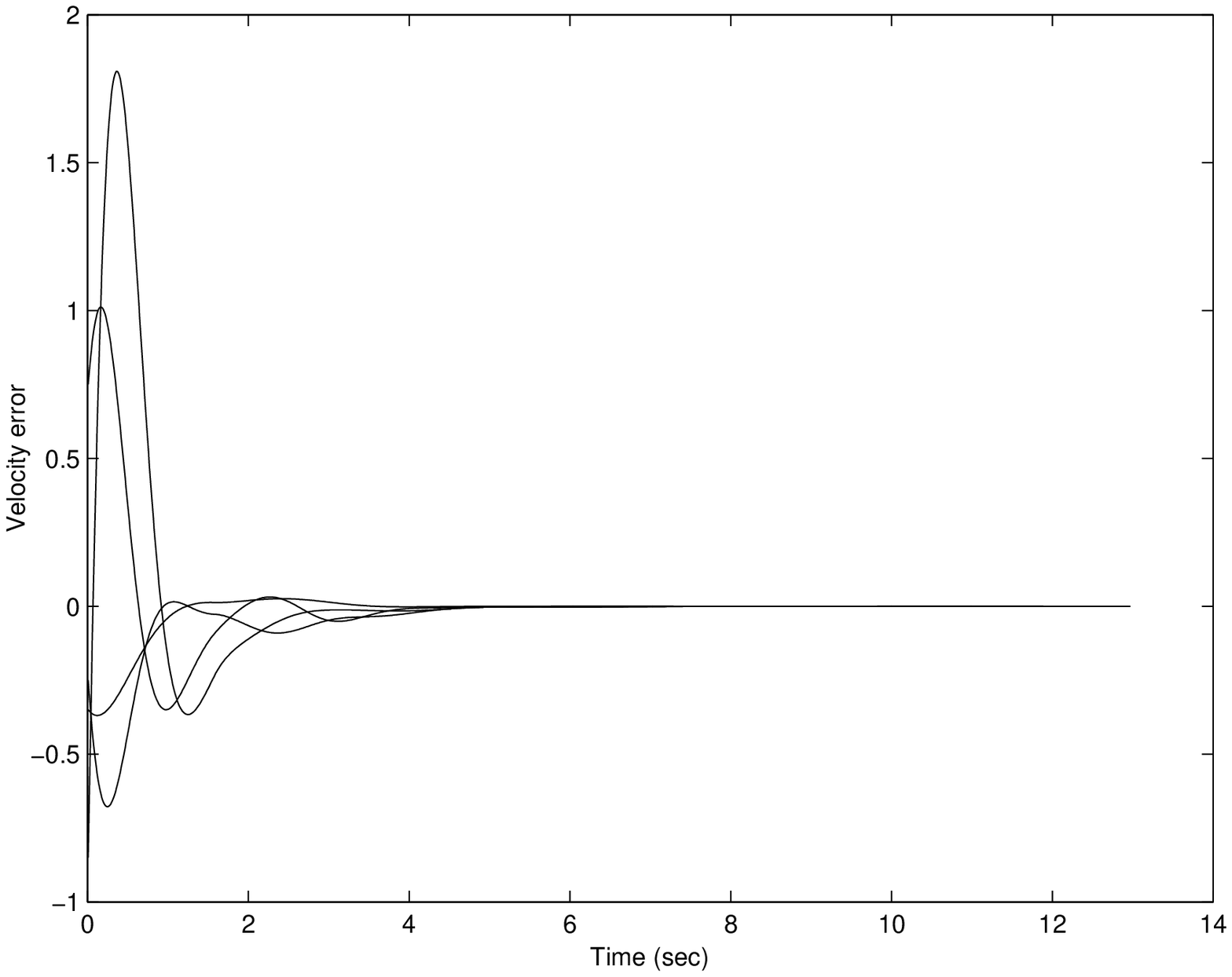}}\\
{\small Fig. 3. Leader-following errors of four agents with the
coupling topology shown in Fig.1}
\end{center}

\begin{center}{\includegraphics[scale=0.4]{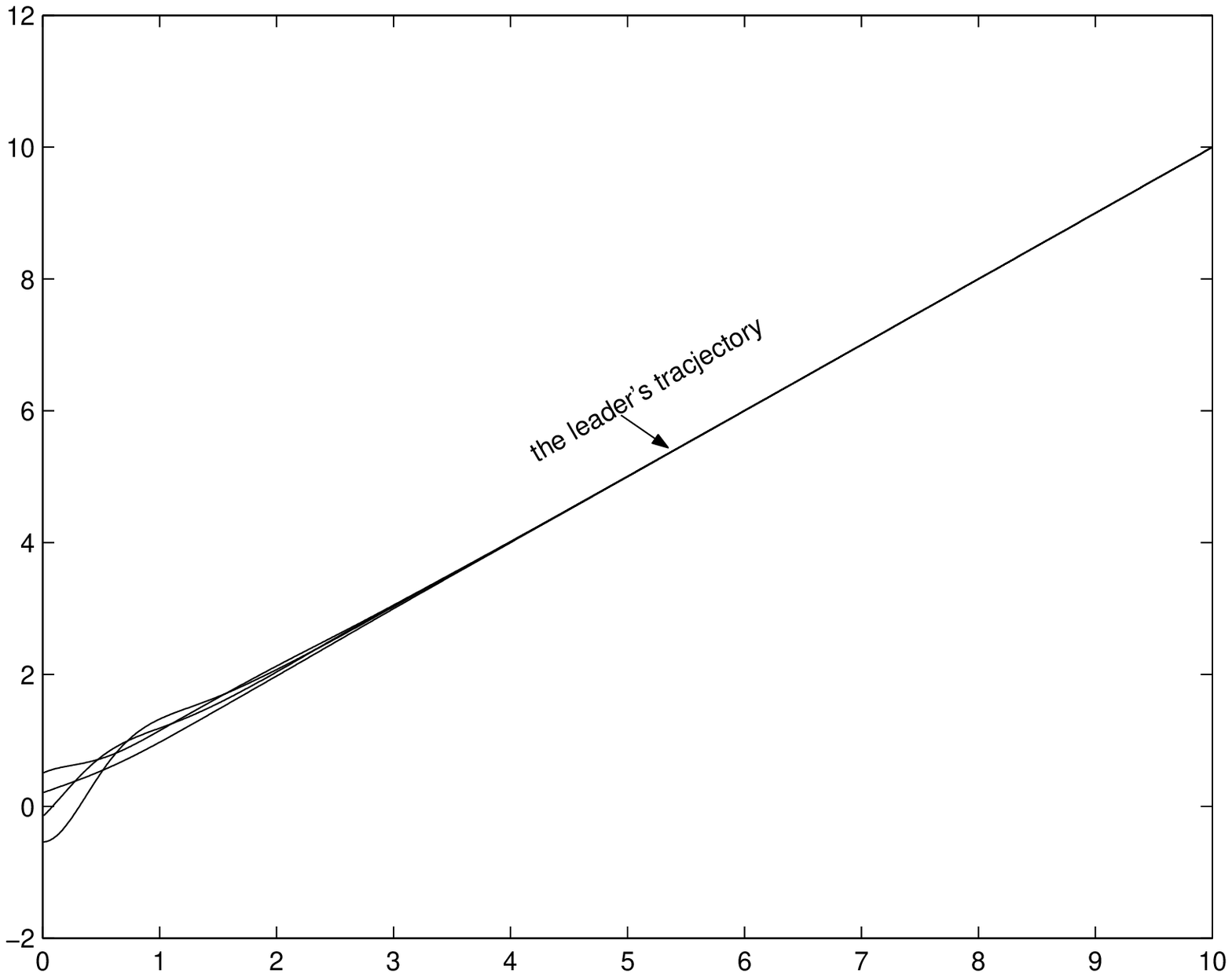}\includegraphics[scale=0.4]{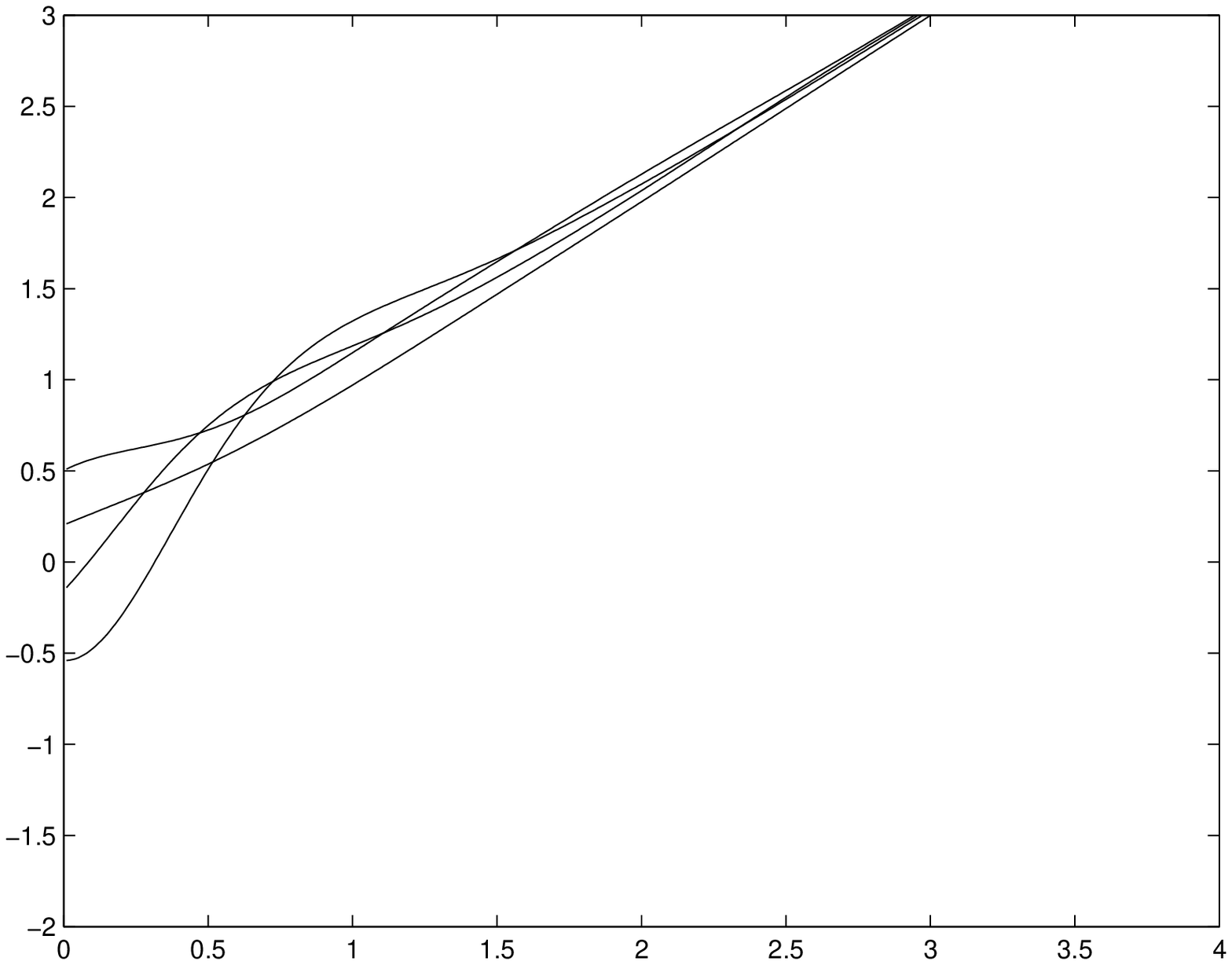}}\\
{\small Fig. 4. Trajectories of four agents and the leader with
the coupling topology shown in Fig.1}
\end{center}

\section{Switched Coupling Topology}

Consider system (\ref{model1}) with switched coupling topology.
Still taking $\bar{x}=x-x_0\mathbf{1},\bar{v}=v-v_0\mathbf{1}$, we
have
\begin{equation}
\label{model4} \dot{\epsilon}=C\epsilon(t)+E_\sigma\epsilon(t-r).
\end{equation}
where $\sigma$ is the switching signal as defined in Section 2,
and
$$
E_\sigma=\begin{pmatrix} 0&0\\
-H_\sigma&0
\end{pmatrix}, \quad H_\sigma=L_\sigma+B_\sigma.
$$

At first, we study the matrix $H_\sigma=L_\sigma+B_\sigma$.

\begin{lemma}
\label{lemgr4} Suppose $\mathcal{G}_\sigma$ is balanced.  Then
$H_\sigma+H_{\sigma}^{T}$ is positive definite if and only if node
$0$ is globally reachable in $\bar {\mathcal{G}}$.
\end{lemma}

Proof: (Necessity) The proof is quite trivial and omitted here.

(Sufficiency)  Because $\mathcal{G}_\sigma$ is balanced, it is
strongly connected if it has a globally reachable node. Then from
Theorem 7 in \cite{olf04}, $\frac{1}{2}(L_\sigma+L_\sigma^T)$ is a
valid Laplacian matrix with single zero eigenvalue. After some
manipulations, it is not difficult to obtain that
$\frac{1}{2}(L_\sigma+L_\sigma^T)+B_\sigma$ is positive definite
(the details can be found in \cite{hong}) and so is
$H_\sigma+H_\sigma^T$.

If ${\cal G}_\sigma$ has no globally reachable node, then there is
no arc between every pair of distinct strong components and we can
renumber the nodes so that Laplacian associated with
$\mathcal{G}_\sigma$ has the form
\begin{equation}
L_\sigma=\left(\begin{array}{cccc} L_{11}(\sigma)&&&\\
&L_{22}(\sigma)&&\\
&&\ddots&\\
&&&L_{pp}(\sigma)
\end{array}\right)
\end{equation}
where each $L_{ii}(\sigma)$ is Laplacian associated with a strong
component $S_i$ for $i=1,...,p,p>1$. By the assumption that node
$0$ is globally reachable in $\bar{\mathcal{G}}_\sigma$, then each
diagonal block matrix $B_i(\sigma)$, corresponding to
$L_{ii}(\sigma)$, is nonzero.  Then, it is easy to see that
$\frac{1}{2}(L_{ii}(\sigma)+L_{ii}(\sigma^T))+B_i(\sigma)$ is
positive definite and therefore, $H_\sigma+H_\sigma^T$ is positive
definite. \hfill\rule{4pt}{8pt}

Based on the balanced graph ${\cal G}_{\sigma}$ (with Lemma
\ref{lemgr4}) and the fact that the set $\mathcal{I}_\Gamma$ is
finite, both $ \tilde{\lambda}=\min\{\mbox{eigenvalues of}\;
H_\sigma+H_\sigma^T\}>0 $ and $
\tilde{\mu}=\max\{\mbox{eigenvalues of}\; H_\sigma H_\sigma^T\}>0$
can be well defined.

\begin{theorem}
\label{thm2} For system (\ref{model4}) with balanced graph ${\cal
G}_{\sigma}$, take
\begin{equation}
\label{kstar1} k>k^*=\frac{\tilde{\mu}}{2\tilde\lambda}+1.
\end{equation}
If node $0$ is globally reachable in $\bar{\mathcal{G}}_{\sigma}$
and $\tau$ is sufficiently small, then
$$
\lim_{t \to \infty}\epsilon(t)=0.
$$
\end{theorem}

Proof:  Take a Lyapunov-Razumikhin function
$V(\epsilon)=\epsilon^{T}\Phi\epsilon$, where
$$
\Phi=\left(\begin{array}{cc}kI_n&I_n\\I_n&I_n
\end{array}\right)\qquad ( k>1)
$$
is positive definite.

Similar to the analysis in the proof of Theorem \ref{thm1}, we can
obtain
$$
\dot{V}\leq \epsilon^{T}(F_\sigma^{T}\Phi+\Phi F_\sigma)
\epsilon+r\epsilon^T\Phi E_\sigma C_\sigma
\Phi^{-1}C_\sigma^TE_\sigma^T\Phi\epsilon+\int_{-r}^{0}
 \epsilon^T(t+s)\Phi\epsilon(t+s)ds.
$$

Take $\phi(s)= q s$ for some constant $q>1$.  In the case of
\begin{equation}
\label{cond31}V(\epsilon(t+\theta))<qV(\epsilon(t)),\quad-\tau\leq
\theta \leq 0,
\end{equation}
we have
$$
\dot{V} \leq -\epsilon^{T}Q_\sigma \epsilon+r\epsilon^T(\Phi
E_\sigma C_\sigma
\Phi^{-1}C_\sigma^TE_\sigma^T\Phi+q\Phi)\epsilon,
$$
where
$$
Q_\sigma=-(F_\sigma^T\Phi+\Phi F_\sigma)=\begin{pmatrix} H_\sigma^T+H_\sigma&H_\sigma^T\\
H_\sigma & 2(k-1)I_n\end{pmatrix}.
$$
$Q_\sigma$ is positive definite for any value of $\sigma$ and then
$\dot V(\epsilon)$ is negative definite if we take (\ref{kstar1})
and
\begin{equation}
\label{bound2} r<\tau=\frac{\lambda_{min}}{\frac{2k}{k-1}\tilde
\mu+\frac{1}{2}q(k+1+\sqrt{(k-1)^2+4})},
\end{equation}
where $\lambda_{min}$ denotes the minimum eigenvalue of all
possible $Q_\sigma$. Thus, the conclusion is obtained according to
Lemma \ref{lem1}. \hfill\rule{4pt}{8pt}

In the switching case, the assumption of balanced graph ${\cal
G}_{\sigma}$ is not necessary for the stability result in Theorem
\ref{thm2}.  The following numerical example shows that the
stability can be obtained even if the coupling topology graph is
not balanced sometimes.

Here we consider there are two coupling topologies, given in Figs.
1 and 2, switching between each other, with the following
switching order:
$\{\bar{\mathcal{G}}_1,\bar{\mathcal{G}}_2,\bar{\mathcal{G}}_1,\bar{\mathcal{G}}_2,...\}$.
With simple calculations, we have
$$
\tilde\lambda=0.5028,\;\tilde\mu=7.9257,\;k^*=7.8816,
$$
$$
\lambda_{min}=0.4781,\;q=1.0500,\;\tau= 0.0174.
$$
Take $k=9$ and the time-varying delay $r(t)=0.0150|\cos(t)|$. Then
the simulation results are shown in Fig. 5.
\begin{center}{\includegraphics[scale=0.4]{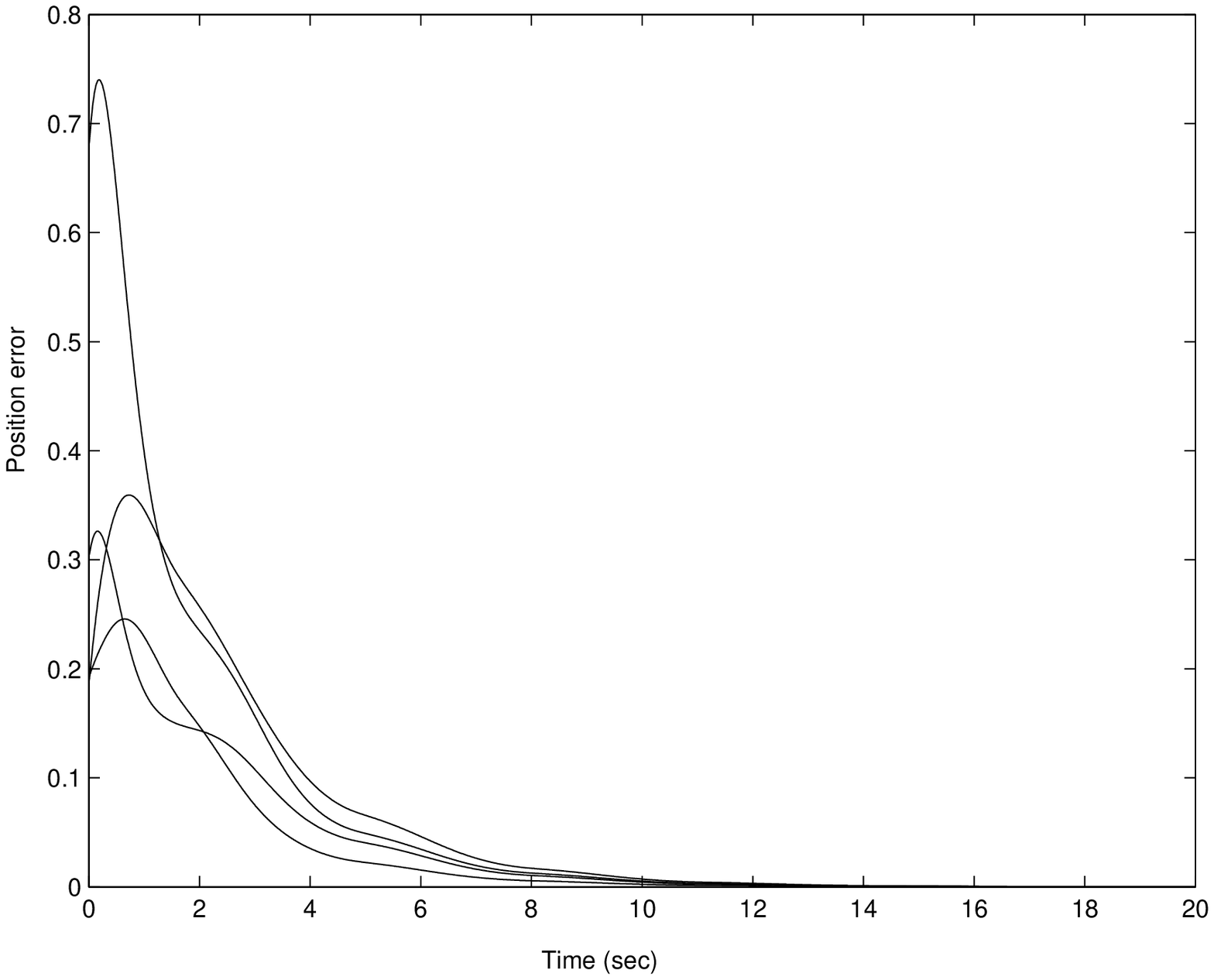}\includegraphics[scale=0.4]{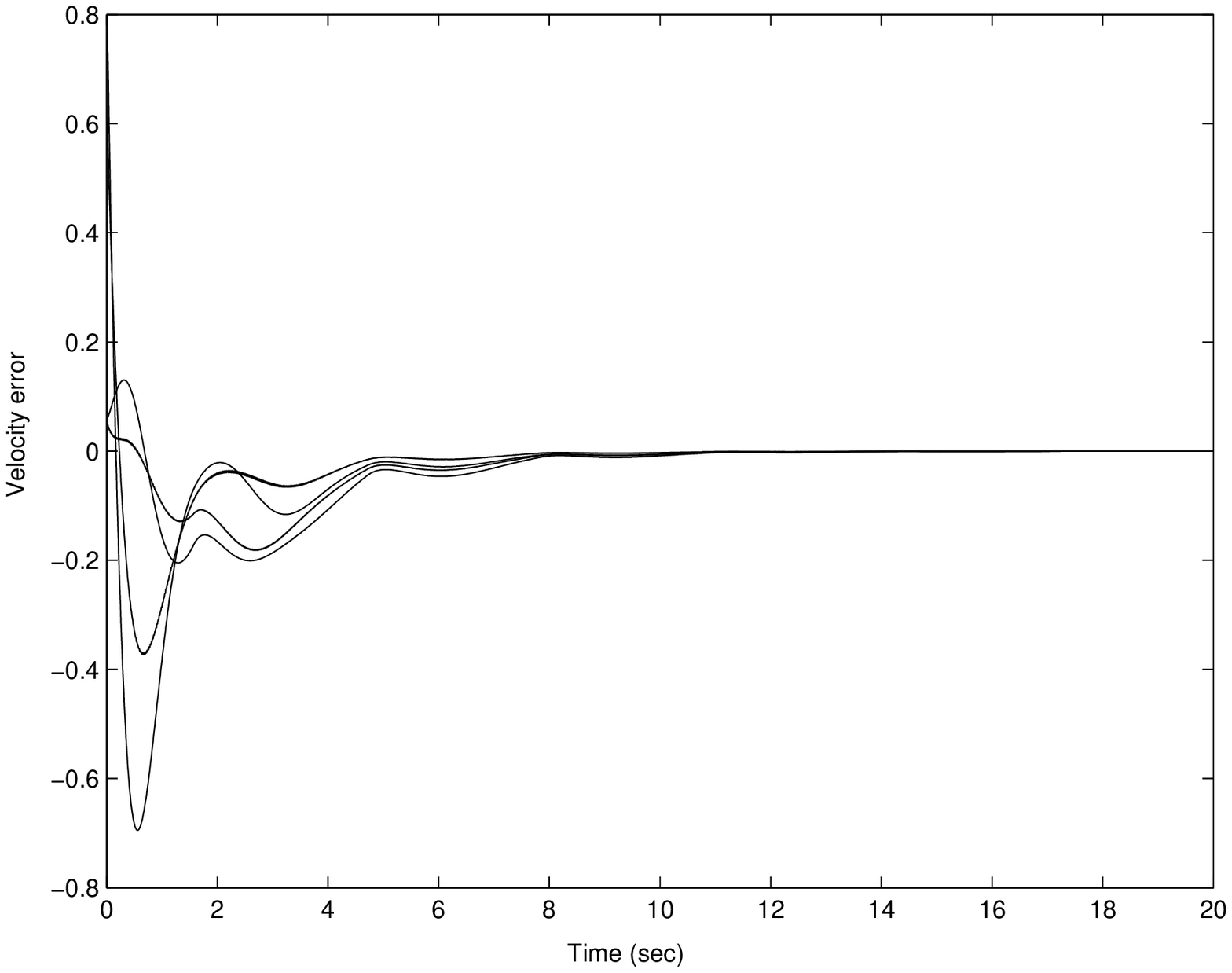}}\\
 {\small Fig. 5. Leader-following errors with two switching graphs given in Fig.1 and Fig.2}
\end{center}

\section{Conclusions}

This paper addressed a coordination problem of a multi-agent
system with a leader.  A leader moves at the constant velocity and
the follower-agents follow it though there are time-varying
coupling delays.   When the coupling topology was fixed and
directed, a necessary and sufficient condition was given.  When
the coupling topology was switched and balanced, a sufficient
condition was presented.  Moreover, several numerical simulations
were shown to verify the theoretical analysis.

\section*{Acknowledgment}

This work was supported by the NNSF of China under Grants
60425307, 50595411, and 60221301.

\end{document}